\documentclass[useAMS, usenatbib]{mn2e}

\usepackage[dvips]{graphicx,color}
\usepackage[]{txfonts}
\usepackage[dvips, bookmarks, colorlinks=false, pdftitle={GRB afterglows from realistic density
profiles}, pdfauthor={P. Mimica and D. Giannios}]{hyperref}

\def\simless{\mathbin{\lower 3pt\hbox
{$\rlap{\raise 5pt\hbox{$\char'074$}}\mathchar"7218$}}}   %< or of order
\def\simmore{\mathbin{\lower 3pt\hbox
{$\rlap{\raise 5pt\hbox{$\char'076$}}\mathchar"7218$}}}   %> or of order

\newcommand{\be}{\begin{equation}}
\newcommand{\ee}{\end{equation}}

\title[GRB afterglows from realistic density
profiles]{GRB afterglow light curves from realistic density
profiles}
\author[P. Mimica and D. Giannios]{P. Mimica$^{1}$\thanks{E-mail:
    Petar.Mimica@uv.es} and D. Giannios$^{2}$\\
$^{1}$Departamento de Astronom\'ia y Astrof\'isica, Universidad de
Valencia, 46100, Burjassot, Spain\\
$^{2}$Department of Astrophysical Sciences, Peyton Hall, Princeton University, Princeton, NJ 08544, USA}

\begin{document}

\maketitle

\label{firstpage}

\begin{abstract}
  The afterglow emission that follows gamma-ray bursts (GRBs) contains
  valuable information about the circumburst medium and, therefore,
  about the GRB progenitor. Theoretical studies of GRB blast waves,
  however, are often limited to simple density profiles for the
  external medium (mostly constant density and power-law $R^{-k}$
  ones). We argue that a large fraction of long-duration GRBs should
  take place in massive stellar clusters where the circumburst medium
  is much more complicated. As a case study, we simulate the
  propagation of a GRB blast wave in a medium shaped by the collision
  of the winds of O and Wolf-Rayet stars, the typical distance of
  which is $d\sim 0.1-1$ pc. Assuming a spherical blast wave, the
  afterglow light curve shows a flattening followed by a shallow break
  on a timescale from hours up to a week after the burst, which is a
  result of the propagation of the blast wave through the shocked wind
  region. If the blast wave is collimated, the jet break may, in some
  cases, become very pronounced with the post break decline of the light
  curve as steep as $t^{-5}$. Inverse Compton scattering of
  ultra-violet photons from the nearby star off energetic electrons in
  the blast wave leads to a bright $\sim$GeV afterglow flare that may
  be detectable by {\it Fermi}.

\end{abstract}

\begin{keywords}
Hydrodynamics -- Shock waves --
gamma-rays: bursts -- radiation mechanisms: nonthermal -- radiative transfer
\end{keywords}

\section{Introduction}

The prompt phase of GRBs is followed by a long-lived afterglow
emission.  The afterglow is believed to be powered by shocks driven by
the relativistic outflow into the circumburst medium. Assuming that
these external shocks inject ultrarelativistic electrons into the
downstream shocked fluid, synchrotron emission from these particles
can account for the basic properties of a large number of the
afterglow observations \citep{Sari:1998kx,Wijers:1999aa}.

Afterglow modeling can provide important information about, among
other things, the density profile of the circumburst material, thus
constraining the nature of the progenitor. So far, however, typically
only very simplistic density profiles have been considered, i.e., the
external medium is assumed either to have constant density or to
follow a power-law function of spherical distance $R$.  Observations
are inconclusive about the density profile, i.e., depending on the
burst and the analysis more than one or none of the simple density
profiles may account for observations (see, e.g.,
\citealt{PK02,Piro05,Starling08,Curran09,Schulze11}).

At least some long-duration GRBs have been convincingly shown to be
associated with the death of massive stars of Wolf-Rayet type
\citep{Galama98,Stanek03,Hjorth03,Mazzali03}. Because Wolf-Rayet stars
are characterized by strong winds, a $R^{-2}$ wind-like profile is, at
first sight, a natural choice to describe the density distribution
surrounding the progenitor. However, for high enough density of the
medium that confines the stellar wind, the termination shock of the
wind may take place sufficiently close to the progenitor to affect the
afterglow light curve \citep{Wijers01,CLF04,RR05,PW06}.

The actual density profile which decelerates the GRB-driven blast wave
could be much more complicated. Massive stars rarely form in
isolation; they preferentially reside in dense stellar clusters where
tens or even hundreds Wolf-Rayet and O stars are crowded on sub-pc
scale regions (e.g. \citealt{MH98}).  About {\it one third} of the
Galactic Wolf-Rayet stars are located in several very massive stellar
clusters (e.g., \citealt{Figer04}).  It is reasonable to expect a fair
fraction of GRBs to take place in such dense stellar environment.  The
interactions of the strong stellar winds in stellar clusters
complicate the medium that surrounds the GRB progenitor and,
therefore, the blast wave evolution.  Furthermore, nearby O stars
provide a strong UV photon field that is up-spattered by electrons
accelerated at the forward shock and may potentially result in bright
GeV afterglow flaring \citep{Giannios:2008ie}.

As a first step towards studying the afterglow appearance in more
realistic density profiles, we focus on a blast wave propagating
through a medium shaped by colliding stellar winds. Hydrodynamic
simulations are used to study the profile from the collision of
stellar winds. Follow-up relativistic hydrodynamic simulations are
performed in order to study the blast wave propagation through such
density profile (Section~\ref{sec:bwcomplex}). The fluid dynamics are
coupled to a radiative transfer code to calculate the resulting
synchrotron and inverse Compton emission (Sections \ref{sec:emission}
and \ref{sec:lcs}). We discuss our results in
Section~\ref{sec:discussion}.

\section{A blast wave in complex density profiles}
\label{sec:bwcomplex}

With the number of massive stars $N_*\sim 100$ 
(ranging from a few tens to a few hundreds) 
crowded on a $R_{\rm c}\simless 1$ pc scale of a
typical massive stellar cluster (similar to, e.g., Westerlund 1,
Arches, Quintuplet, or Center in the Galaxy), the mean distance between O stars is
$d\sim R_{\rm c}/N_*^{1/3}\sim 0.1$ pc. The blast wave driven by a GRB, 
still relativistic at these distances, is expected to encounter
density bumps while propagating in the cluster. 
We simulate several density profiles that may be expected in a massive
cluster and then study the blast wave propagation through them.

\subsection{Colliding winds}

%%%
\begin{figure}
\includegraphics[width=8.5cm]{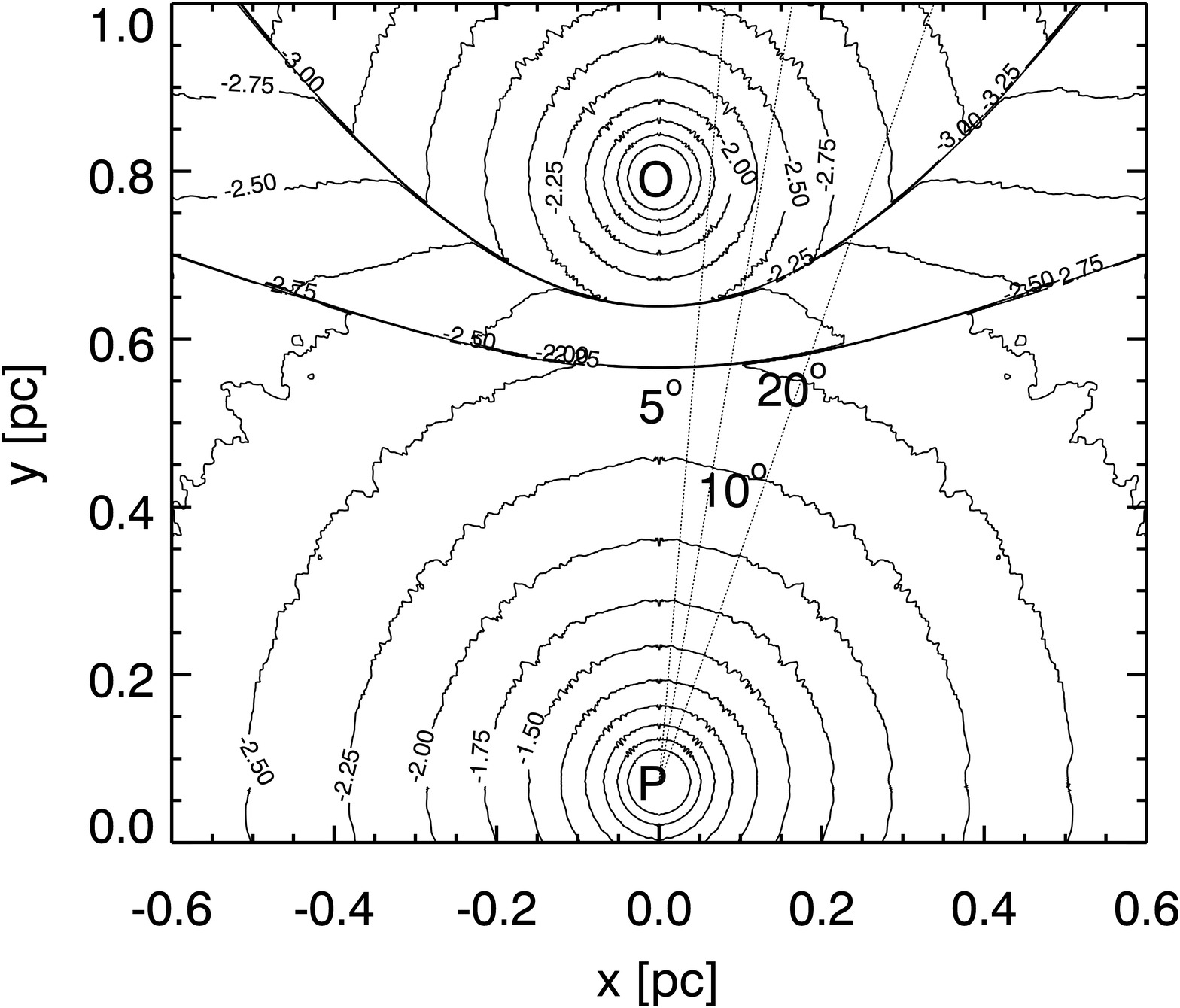}
\caption{Contours of the logarithm of the gas density (in arbitrary
  units) in the vicinity of the progenitor (P) and the O star. The
  contour values are decreasing in steps of $0.25$ as one moves away
  from the stars. The shocked stellar winds are located in the region between two thick arcs,
  which are the termination shocks for the progenitor and O star
  winds. The three lines show the cuts at angles $5$, $10$ and
  $20$ degrees from the line joining the centers of the two stars,
  respectively.}
\label{fig:winds}
\end{figure}
%%%

In the young stellar cluster under consideration, the gas density is
shaped by the interactions of the stellar winds. Since the most
massive (O and Wolf-Rayet) stars have the strongest winds, they are
going to dominate these interactions. The stars are surrounded by
regions of their freely expanding winds that are followed, further
out, by regions of shocked gas-result of wind-wind collisions. For a
given line of sight from the explosion center to the observer, the
density is likely to be shaped by the few massive stars that happen to
lie close to it. For $N_*$ stars randomly distributed in the cluster,
the closest one to the line of sight of the observer is located at an
angle $\theta\sim (4/N_*)^{1/2}\sim 0.2N_{*,2}^{-1/2}$ rad (where the
$A=10^xA_X$ notation is adopted). Encountering a star within an angle
$\theta\sim 10^{\rm o}$ from the line of sight is, therefore, the norm
in the massive clusters under consideration. As long as the outflow is
ultrarelativistic with a bulk Lorentz factor $\Gamma\gg 1$, the
observer of the GRB afterglow probes the blast-medium interactions
that take place within a narrow cone of angle $\sim 1/\Gamma$ with
respect to the line of sight. In the narrow cone which the observer
can see the closest stellar encounter is the dominant one in
determining the relevant density profile.

As a first approach, we limit ourselves to a single massive star ($O$)
located at distance $d \simless 1$ pc from the explosion center ($P$)
and at a modest angle $\theta\simless 0.5$ rad with respect to the
line that connects the explosion center to the observer
(Fig.~\ref{fig:winds}). 
We consider the density profile resulting from the collision of the
winds of two stars with mass loss rates $\dot{M}_{\rm
    P}=10^{-5}M_\odot yr^{-1}$ and $\dot{M}_{\rm O}=10^{-6}M_\odot
yr^{-1}$, respectively. The $\dot{M}_{\rm P}$ is typical for a
Wolf-Rayet star assumed to be the GRB progenitor, and $\dot{M}_{\rm
    O}$ is expected for a typical O star (also
referred to as the `companion' star). The stellar winds are assumed
to be cold and have the same velocity $v_{\rm w}=1000$ km s$^{-1}$. We have
performed a 2-dimensional (2D) axisymmetric hydrodynamical simulation
of such wind-wind interaction assuming adiabatic behaviour of the
gas\footnote{\citet{Stevens:1992kx} and, more recently,
  \citet{vanMarle:2011jt} have studied the stellar wind interaction in
  detail and showed that for the distances $d\simmore 0.1$ pc of
  interest here, radiative cooling in the shocked regions is
  negligible, thus justifying the assumption that the gas is
  adiabatic.}. We have used the high-resolution shock-capturing scheme
\emph{MRGENESIS} \citep{Mimica:2007aa,Mimica:2009bb} to run the 
simulation until a steady state is established. The numerical resolution is 800 and 2400 cells in $x$ and
$y$ directions, respectively. 

In Fig.~\ref{fig:winds}, we show the density contours from such
interaction. The regions of freely
expanding winds and shocked winds are clearly seen. Because of the
choice of the same velocity $v_{\rm w}$ for the two winds, there is no
contact discontinuity separating the two shocked winds
\citep[see][]{Stevens:1992kx}. The shocked wind region lies closer
(and bends toward to) the O star because $\dot{M}_{\rm P}>\dot{M}_{\rm O}$.

%%%
\begin{figure}
\includegraphics[width=8.5cm]{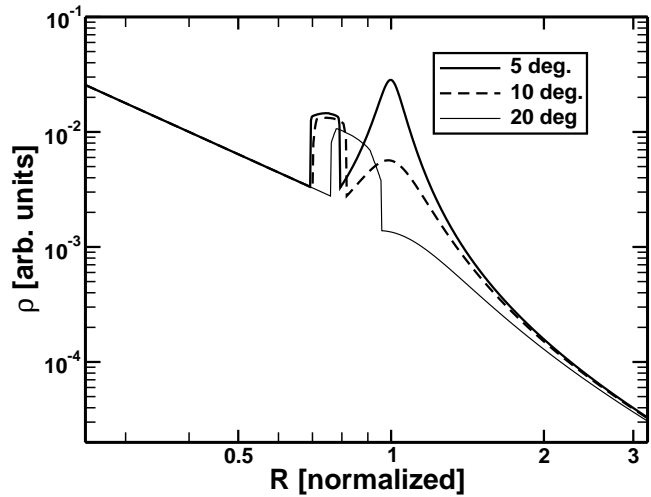}
\caption{Density profiles along the cuts at $\theta=5$, $10$ and $20$
  degrees from the line joining the centers of the two stars on
  Fig.~\ref{fig:winds}. The radius $R$ is measured from the center of
  the progenitor and is normalized to the distance between the stars.
}
\label{fig:profiles}
\end{figure}
%%%

In Fig.~\ref{fig:profiles}, we plot the density profile moving
radially outwards from the Wolf-Rayet star and for different angles
with respect to the O star.  The density profiles are averaged over a
cone with opening angle of $5$ degrees. One can see the $R^{-2}$
density profile from the progenitor, a narrow region of enhanced
density due to the shocked colliding winds and, at larger distance the
freely expanding wind of the companion.  These profiles are used as
the density of the external medium through which the GRB blast wave
propagates.

\subsection{Blast wave propagation}
\label{sec:bw}

For characteristic distance $d\simmore 10^{17}$ cm between the stars
considered here, the blast wave has evolved into a self-similar
\citep{Blandford:1976yg} stage before encountering the shocked wind
region. As long as the bulk Lorentz factor of the shocked fluid is
larger than the inverse opening angle of the jet: $\Gamma\simmore
\theta_j^{-1}$, 2D effects (e.g. lateral jet
spreading) can be safely neglected. In this regime, the blast wave can be
studied assuming a spherically symmetric evolution. We follow the
evolution of the blast wave with 1D relativistic hydrodynamic
simulations \citep{Mimica:2009bb}. We caution, however, that by the
end of our simulations, the fluid may decelerate to $\Gamma\sim$
several and lateral spreading effects (not taken into account by our
simulations) may be modestly important for $\theta_j\sim 0.1$; a value commonly
inferred in GRBs (Frail et al. 2001). This point is discussed further
in Section 3.1.

\begin{figure}
\includegraphics[width=8.5cm]{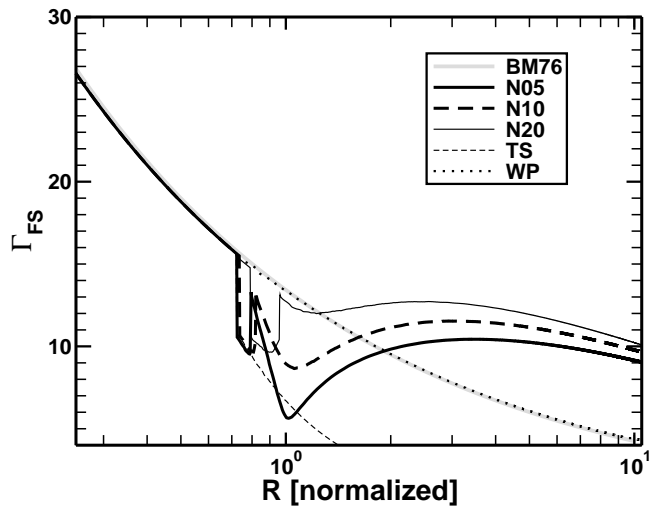}
\caption{Lorentz factor of the shocked fluid behind the forward shock as a
  function of the distance from the progenitor for a blast wave
  propagating at an angle $\theta=5$, $10$ and $20$ degrees (thick
  full, thick dashed and thin full lines, respectively) from the line
  joining centers of the two stars (see also
  Fig.~\ref{fig:winds}). Also shown is the evolution of the Lorentz
  factor in a wind profile (dotted line), the \citet{Blandford:1976yg}
  analytic solution for a wind profile (thick gray line closely
  following the dotted line) and the evolution in a wind profile with
  the termination shock (thin dashed line).  The distance is
  normalized to the separation of the two stars. }
\label{fig:gammafs}
\end{figure}

Having specified the density profile for the external medium, we only
need to choose the (isotropic equivalent) energy of the blast $E$ for
the self-similar initial conditions to be well defined.  We consider 
a rather powerful GRB of $E=10^{54}$ erg. The distance between the 
stars is set to $d=2.2 \times 10^{18}$ cm. The blast wave dynamics is followed
with relativistic hydrodynamic simulations using the code 
\emph{MRGENESIS} described in
\citet{Mimica:2009bb}. The simulations have been performed in
spherical symmetry with a numerical resolution of $16000$ cells in the
blast wave. We have generated five blast wave models:
\begin{itemize}
  \item three models with external medium density profiles shown in
    Fig.~\ref{fig:profiles}: $N05$ (cut along a line $5$ degrees off
    the line joining centers of two stars), $N10$ ($10$ degrees) and
    $N20$ ($20$ degrees)
  \item a simulation in an external medium with a wind profile
    throughout: model $WP$
  \item a simulation in an external medium with a wind profile out to the
    wind termination shock (TS) at $1.6\times 10^{18}$ cm (we assume a
    density jump of factor $4$ at the TS and a constant density medium
    afterwards): model $TS$
\end{itemize}
In Fig.~\ref{fig:gammafs}, we show the Lorentz factor of the fluid at
the forward shock as function of distance from the center of the
explosion. The initial evolution of all models is that of a self-similar
blast in a $R^{-2}$ wind profile where the Lorentz factor just behind
the forward shock $\Gamma\propto R^{-1/2}$
\citep{Blandford:1976yg}. For models $N05$, $N10$ and $N20$ the 
self-similar evolution holds until distance $R\sim d$ where the blast
encounters the region of the shocked winds. At this stage, the
enhancement of the density causes the forward shock to decelerate and
a weak reverse shock launches into the blast. On exit from the shocked
wind region, the density drops and the blast rarefies in the
front. The density profile has a second bump due to the approach to
the companion star. This results in one more drop in the Lorentz
factor $\Gamma$, the depth of which depends on the angle
$\theta$. These non self-similar stages of the interaction can only be
studied in detail using numerical simulations. At still larger
distance the blast wave relaxes to a self-similar solution determined
by the $R^{-2}$ density profile of the companion star. In the model
$WP$ the blast wave follows the \citet{Blandford:1976yg} evolution,
while in the model $TS$ it decelerates much faster after encountering
the constant density medium behind the termination shock (e.g. van
Eerten et al. 2009).

%%%
%%%

\section{Emission}
\label{sec:emission}

After the hydrodynamical simulations have been performed we use the
\emph{SPEV} code \citep{Mimica:2009aa} to compute the light
curves. The details of how \emph{SPEV} is applied to calculate the
afterglows from GRBs can be found in the section 3 of
\citet{Mimica:2010cc}. We consider synchrotron emission from the
shock-accelerated electrons (Sect. 3.1) and external inverse Compton
scattering of the photon field from the companion star off the same
electrons (Sect. 3.2). In this paper we ignore the synchrotron self
Compton process. Sect. 3 contains qualitative discussion and 
analytical estimates while the numerical results are presented in
Sect. 4.

\subsection{Synchrotron emission}

During the initial self-similar stage of deceleration in the wind of the
progenitor of density $\rho=A/R^2$, with $A\equiv \dot{M}_{\rm P}/4\pi
v_{\rm w}=5\times 10^{11}A_*$ gr$\cdot$cm$^{-1}$ where $A_*=\dot{M}_{-5}v_{\rm
  w,8}^{-1}$, the bulk Lorentz factor of the fluid just behind the shock is
\be \label{eq:gamma}
\Gamma=\sqrt{\frac{9E}{16\pi ARc^2}}=60
E_{54}^{1/2}R_{17}^{-1/2}A_*^{-1/2}.  
\ee
The observer time evolves as $t_{\rm obs}=R/2\Gamma^2c\simeq
500R_{17}^2A_*E_{54}^{-1}$ s (neglecting cosmological redshift
effects). For the distance $d\sim 3\times
10^{17}$ cm and the energy $E\sim 10^{53}$ erg, it takes typically a
day for the blast wave to reach the shocked wind region (this
timescale can range from hours to a week, depending on the various
parameters).

\begin{figure}
\includegraphics[width=8.5cm]{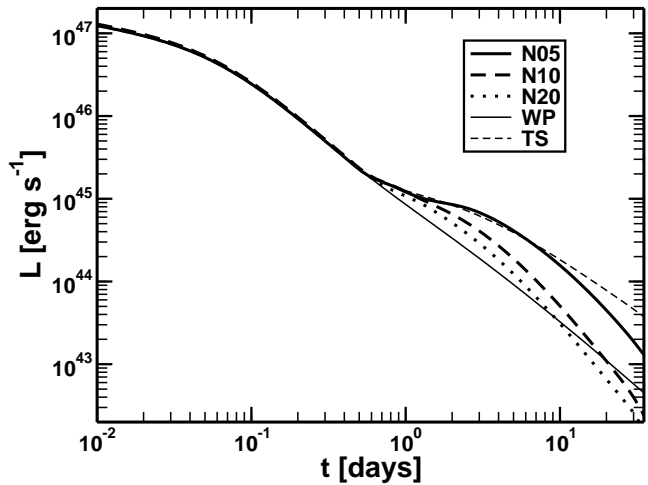}
\caption{R-band ($\nu=4.3 \times 10^{14}$ Hz) light curves for the
  five numerical models. Their light curves are indistinguishable
  until $\simeq 0.5$ days, at which point the light curve of all the
  models which contain a shock (i.e., $N05$, $N10$, $N20$ and $TS$)
  flattens. The $TS$ light curve soon reaches the asymptotic behavior
  expected of the blast wave propagating into a homogeneous external
  medium \citep[see e.g., Fig.~2 of][]{vanEerten:2009em}. The light
  curves of the remaining three models show a more complex behaviour
  due to the crossing over a second wind termination shock. The crossing
  point depends on the angle at which the blast wave is propagating,
  which is reflected in the different late-time light curves for
  $N05$, $N10$ and $N20$ models.}
\label{fig:opt}
\end{figure}

We make standard assumptions in calculating the synchrotron emission
from shocked fluid in the blast wave following \citet{Sari:1998kx},
i.e., we assume that a fraction $\epsilon_e$, and $\epsilon_B$ of the
dissipated energy goes into accelerating electrons into a power-law
distribution with index $p$ and amplifying magnetic field,
respectively. Throughout this paper we fix $\epsilon_e=0.1$,
$\epsilon_B=0.005$, and $p=2.5$. The light curve at the initial 
self-similar stage is that calculated by \citet{Chevalier:2000aa} that
focus on a blast propagating in a $R^{-2}$ wind profile. At a time
$t_{\rm obs}\sim d/\Gamma^2c$ the blast wave encounters the region of
the shocked winds, which encounter causes a flattening in the light
curve (Fig.~\ref{fig:opt}). This feature has already been discussed in
the case of a wind with a termination shock
\citep{Nakar:2007yq,vanEerten:2009em}. In our setup, the blast wave
crosses the shocked wind region on a short timescale and transits to a
less dense wind of the companion. This transition, a distinct
characteristic of colliding stellar winds, leads to a steeper decline
of the light curve.  More light curves and a discussion are presented in
the next section.

In our 1D, spherically symmetric approximation, even the sharp features
of the external medium (e.g. shocks, density jumps) result in smooth
changes on the afterglow emission. This effect, result (at least in
part) of the large lateral extent of the emitting region, has been
studied in detail by \citet{Nakar:2007yq} and
\citet{vanEerten:2009em}. However, there is good evidence that GRB jets are
collimated with opening angles of $\theta_j\sim 0.1$
(e.g. \citealt{Frail01}). Here we show that departures from spherical
symmetry combined with structured external media introduce interesting
novel features to the afterglow light curves.

In addition to modulations of the light curve because of external
density profile, deviations from spherical symmetry are also expected
to affect the afterglow appearance. A `jet break' in the light curve
is expected to occur when $\Gamma\sim \theta_j^{-1}$ for a smooth
density profile \citep{Rhoads:1999rt}.  \citet{Zhang:2009fl} have shown that the main
effect of the $\Gamma\simless \theta_j^{-1}$ transition is a steepening in the light curve because of
the `missing surface' emitting towards the observer (see, however
\citealt{Wygoda:2011wm}). With our 1D simulations we
cannot treat the transition exactly. However, we can easily include
the dominant geometric effect that contributes to the jet break 
by considering the emission taking
place only within an angle $\theta_j$ with respect to the observer. Such
examples are shown in Fig.~\ref{fig:opt-breaks} and discussed in
Section 4.1.

\subsection{Compton scattering of the photons from the companion}
\label{sec:compton}

When the blast wave approaches the companion, it is exposed to the
dense UV field of the O star. Electrons accelerated at the shock will
Compton up-scatter the stellar light to (typically) GeV energies. Such
encounter leads to a GeV flare hours to days after the GRB (the time
of closer approach to the star) that is potentially detectable with
{\it Fermi}. \citet{Giannios:2008ie} analytically derived the
$\gamma$-ray fluence from a blast wave sweeping past the photon field
of an O star for constant density ISM. Here we repeat the derivation
for the wind profile, so that a comparison of the analytical estimates
to more accurate numerical results can be done
(Section~\ref{sec:he}). In addition to the model of
\citet{Giannios:2008ie}, we consider the contribution of the ambient
UV photon field of the cluster to the external inverse Compton (EIC)
scattering of blast wave electrons.

We consider an O star of $L_*\sim 10^6L_{\odot}\sim 10^{39.5}L_{39.5}$
erg$\cdot$s$^{-1}$ located at a distance $d$ from the progenitor, and
at an angle $\theta$ with respect to the line of sight. The emission
of the star is assumed to peak at $e_*=10e_{*,1}$ eV (typical for an O
star).  Electrons accelerated at the forward shock have a lower
cut-off of their distribution at $\gamma_{\rm min}\simeq
(\epsilon_e/3)\Gamma (m_p/m_e)\simeq
1000\epsilon_{e,-1}E_{54}^{1/2}d_{18}^{-1/2}A_*^{-1/2}$, where
Eq.~\ref{eq:gamma} is used in the last step. The energy of the
scattered photons in the central engine frame is \be \label{eq:energy}
e_{\rm ic}\simeq 2\Gamma^2\gamma_{\rm min}^2e_*=8
\frac{\epsilon_{e,-1}^2E_{54}^2e_{*,1}} {d_{18}^2A_{*}^2}\quad \rm
GeV.  \ee The detailed numerical results (Section~\ref{sec:he}) identify the
expression $e_{\rm ic}$ as a good estimate for the peak of the
$L_{\nu}$ spectrum of the EIC component.  Note that for $e_{\rm
  ic}\simmore (m_ec^2)^2/e_*\simeq 25/e_{*,1}$ GeV, the scattering
takes place in the Klein-Nishina regime for all electrons and the last
expression is not applicable.

The fluence of the EIC component can be estimated by considering the
fraction of the total energy carried by electrons $\epsilon_e E$ that
is radiated away because of EIC. The closest approach to the companion
star is $\sim \theta d$ with the energy density of photons (in the
rest frame of the blast wave) being $U_{\rm ph}\simeq \Gamma^2L_*/4\pi
(\theta d)^2c$. The IC {\it cooling} timescale of an electron with
$\gamma_{\rm min}$ is $t_{\rm cool}=2\times 10^7/\gamma_{\rm
  min}U_{\rm ph}$ s. The energy density of external photons peaks
while the blast travels distance $d\theta/\Gamma$ (in the comoving
frame of the blast).  The {\it residence} time in the intense
radiation field is, therefore, $t_{\rm res}=r\theta/\Gamma
c$. Combining all these, we obtain the fluence of the IC component:
\be \label{eq:fluence} F_{\rm ic}=\frac{t_{\rm res}}{t_{\rm
    cool}}\epsilon_e E=5\times
10^{50}\frac{\epsilon_{e,1}^2E_{54}^2L_{39.5}}{\theta_{-1}d_{18}^2A_*}\quad
\rm erg.  \ee The detailed numerical results (see
Section~\ref{sec:he}) show that expression for $F_{\rm ic}$ {\it
  overestimates} the fluence of the EIC component by a factor of,
typically, $\sim$ a few. Klein-Nishina corrections contribute mostly
to the discrepancy.

The flare peaks at time \be \label{eq:flarepeak} t_{\rm
  peak}=d/2\Gamma^2c\simeq 0.5 d_{18}^2A_*E_{54}^{-1} \quad \rm days.
\ee Note that for $d\sim 3\times 10^{17}$ cm, $A_*\simless 1$, a close
encounter with a bright star leads to extraction of a large fraction
of the electron energy in the blast through EIC scattering. The
emission from such encounter may peak several hours after the burst.

%%%
%%%

\section{Light curves}
\label{sec:lcs}

In this section we present the light curves produced by the five
models introduced in the section~\ref{sec:bw}. We first discuss the
optical, X-ray light curves result of synchrotron emission, and then 
turn our attention to the $\gamma$-ray emission result of EIC.

\subsection{Optical/X-ray emission}
\label{sec:optical}

Fig.~\ref{fig:opt} shows the optical light
curves. Before reaching the progenitor wind termination
shock the light curves are indistinguishable from the one
corresponding to a self-similar blast wave propagating in a $R^{-2}$
profile (model $WP$). From that point on the models begin to diverge
depending on whether the blast wave continues propagating into a
constant-density environment ($TS$) or whether it eventually
encounters the wind of the companion star ($N05$, $N10$ and
$N20$). The latter three models differ in their light curves because
the blast wave crosses the wind interaction zone and a companion
stellar wind at different angles.

In model $N05$ the blast wave propagates closest to the companion star
and it encounters the densest companion wind. Therefore it is slowed
down more than $N10$, which in turn decelerates more than $N20$ (see
also Fig.~\ref{fig:gammafs}). This is seen in Fig.~\ref{fig:opt} after
$t\simeq 3$ days, where $N05$ is the brightest of the three models,
followed by $N10$ and $N20$.

%%%
\begin{figure}
\includegraphics[width=8.5cm]{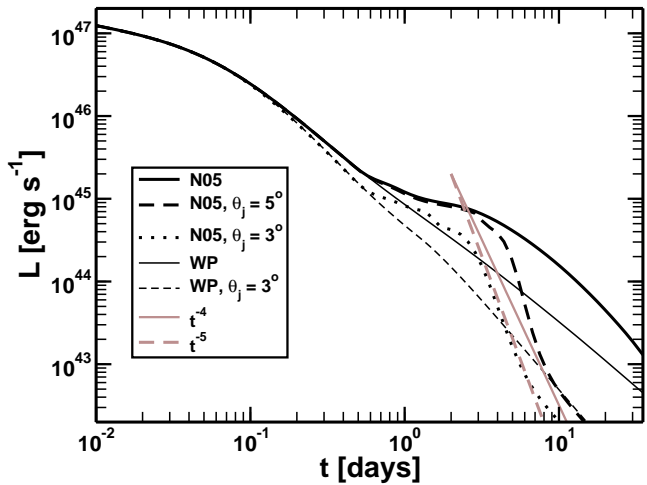}
\caption{Same as Fig.~\ref{fig:opt}, but for models where various jet
  opening angles have been assumed. Shown are the light curves of the
  $N05$ model assuming a spherical blast-wave (thick line), and
  assuming a jet half-opening angle of $5^{\rm o}$ and $3^{\rm o}$ (thick dashed
  and dotted lines, respectively). For comparison we show the light
  curve in a wind environment for the spherical outflow and a jet
  half-opening angle of $3^{\rm o}$ (thin full and dashed lines,
  respectively), and two curves showing the decline proportional to
  $t^{-4}$ and $t^{-5}$ (gray full and dashed lines, respectively).}
\label{fig:opt-breaks}
\end{figure}
%%%

Fig.~\ref{fig:opt-breaks} shows the effect of the finite opening angle
of the jet. Since we are simulating a one-dimensional spherical blast
wave, we model a jet with a half-opening angle $\theta_j$ by assuming
no contribution to emission from the fluid at angle with respect to
the line of sight $>\theta_j$. As expected, while $\Gamma\simmore
\theta_{j}^{-1}$ the jet collimation is not affecting the light
curve. For a blast wave propagating in a pure wind-like profile, a
rather smooth break appears in the light curve when $\Gamma\sim
\theta_{j}^{-1}$. The break can be much sharper if the transition to
$\Gamma< \theta_{j}^{-1}$ takes place when the blast wave reaches the
shocked wind region. The rapid decline of the Lorentz factor of the
blast wave at the density jump is not compensated by the increase of
the emitting surface visible to the observer and the flux drops much
faster than expected from a jet break in a smooth external
medium. Comparing with the thick full and dashed gray lines (which
show the temporal decline proportional to $t^{-4}$ and $t^{-5}$,
respectively), we see that the combination of the jet break and the
interaction of the blast wave with a shocked wind environment can
produce steep declines.

Fig.~\ref{fig:opt-breaks} shows also demonstrates how the break 
happens at progressively earlier times as $\theta_j$ is decreases 
(thick full, dashed and dotted lines, respectively). At later times the light curve decline
becomes less steep due to the acceleration of the blast wave as it leaves the
shocked region and encounters a less dense companion stellar
wind.

\begin{figure}
\includegraphics[width=8.5cm]{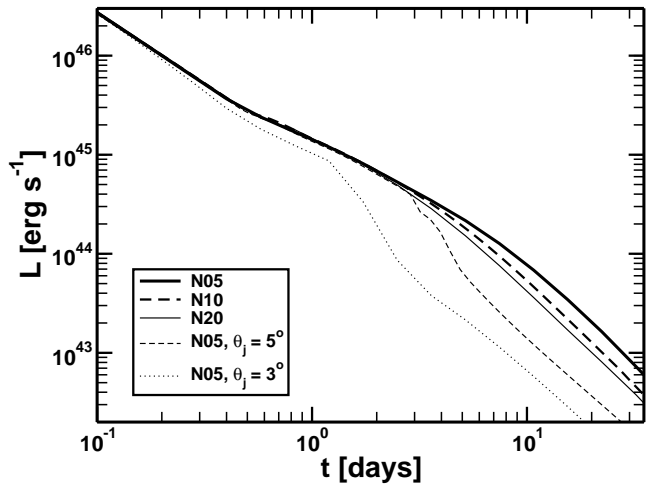}
\caption{$1$ keV X-ray light curve for the models $N05$, $N10$
  and $N20$ (thick full, dashed and dotted lines, respectively). Shown
  are also two cases for the model $N05$ if we assume a jet half-opening
  angle of $5^{\rm o}$ and $3^{\rm o}$ (thin dashed and dotted lines,
  respectively).}
\label{fig:X-ray}
\end{figure}

Fig.~\ref{fig:X-ray} shows the X-ray light curve for the models $N05$,
$N10$ and $N20$, as well as for the model $N05$ assuming a small jet
opening angle, to simulate the effect of a jet break. As can be seen,
the effect of the traversal of the shocked wind leaves qualitatively
similar but less pronounced imprints on the X-rays in comparison to
the optical light curve (see Fig.~\ref{fig:opt}).  The effect of the
jet break (result of a finite jet angle) is, as expected, also an
adiabatic one (e.g., not due to electron cooling), appearing in
simultaneously in the optical and X-rays.

%%%
%%%

\subsection{$\gamma$-ray emission}
\label{sec:he}

The analytical estimates of Section~\ref{sec:compton} provide the
dependence of the photon energy and fluence ($e_{\rm ic}$ and $F_{\rm
  ic}$, respectively) of the EIC emission on the various parameters.
The actual numerical values of Eqs.~\ref{eq:energy} and
\ref{eq:fluence} are, however, meant more as rough order-of-magnitude
estimates then the accurate predictions. In this section we use the
numerical simulations to calculate the EIC emission and calibrate the
analytics.

Part of the uncertainty of the analytical estimates is connected to
the fact that they ignore the blast wave hydrodynamics of the shocked
wind regions (instead, the blast wave is assumed to propagate in a
freely propagating stellar wind). The numerical simulations are more
accurate since they include the effect of the colliding winds on the
blast dynamics and the exact calculation of the EIC cooling during the
blast encounter (including Klein-Nishina effects) for a power-law
injected electron distribution. We compute the EIC emission assuming a
monochromatic external point-source of radiation (good approximation
for the black body stellar emission). To compute the emissivity we use
the method described in the section~2.2.1 of
\citet{Mimica:2004zz}.\footnote {This enables us to use the same
  piecewise power-law representation of the non-thermal electron
  distribution that \emph{SPEV} uses to compute synchrotron emission,
  resulting in a considerable speed-up of the EIC emissivity
  calculation.}

\begin{figure}
\includegraphics[width=8.5cm]{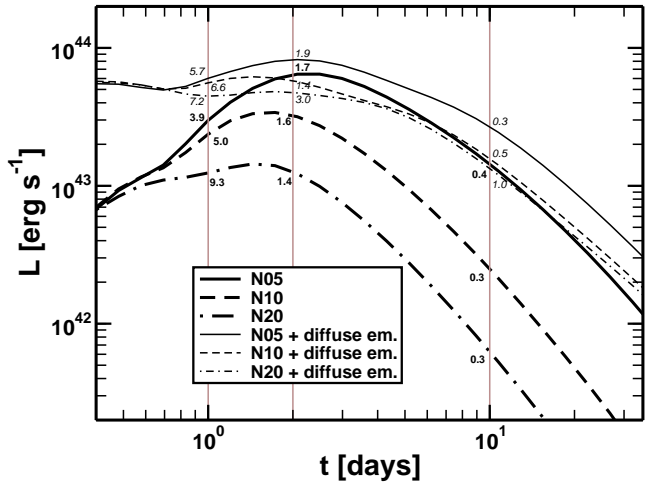}
\caption{Bolometric ($4$ MeV - $4$ TeV) light curves from external
  inverse Compton scattering for the models $N05$, $N10$ and $N20$
  (full, dashed and dotted lines, respectively). The thick lines show
  the light curves assuming that the only source of photons is the
  companion star. The numbers in bold show the energy of the spectral
  peak in GeV for radiation observed $1$, $2$, and $10$ days after the
  burst. The thin lines show the emission when the diffuse cluster
  emission is taken into account as a soft-photon source 
  (see text for details). The numbers
  in italic denote the spectral peak in GeV for this case.}
\label{fig:EC-all}
\end{figure}

Fig.~\ref{fig:EC-all} shows the EIC bolometric light curves and the
energy of the spectral peak (in GeV) for the models $N05$, $N10$ and
$N20$ (thick full, dashed and dotted lines, respectively). The EIC
emission peaks at time $t_{\rm peak}\sim 2$ days in good agreement to
Eq.~\ref{eq:flarepeak} (using the value $d_{18}=2.2$) and then
gradually declines.  The duration of the high energy flare is $\delta
t\sim t_{\rm peak}$. As expected, the EIC is brighter for the closer
encounters (smaller $\theta$) between the line of sight and the
companion star.  At maximum of the EIC luminosity the peak of the
$L_{\nu}$ spectrum is $E\simeq 2$ GeV independently of $\theta$ in
close agreement to Eq.~\ref{eq:energy}. The spectral peak moves to
lower energies as function of time. We have verified that The EIC
spectrum has a cutoff at $\sim 25$ GeV because of the Klein-Nishina
suppression.  The numerically calculated fluence is found to be a
factor of $\sim 3$ less that the estimate in Eq.~\ref{eq:fluence}.

%%%
%%%

The total number of $\gamma$-ray photons emitted in the 0.1-300 GeV
energy range (where the effective area of {\it Fermi-LAT} peaks) is
$N_{\gamma}\simeq 8\times 10^{51}$, $2\times 10^{51}$, $6\times 10^{50}$ for
$\theta=5^{\rm o},10^{\rm o}, 20^{\rm o}$, respectively. For an
effective area of {\it LAT} of $A_{\rm eff}\sim 10^4$ cm$^2$, the closest encounter
can be detected out to a proper distance of $d_{\rm p}\simeq
(A_{eff}N_{\gamma}/4\pi)^{1/2}\simeq 2.5\times 10^{27}$ cm or out to a
modest redshift of $z\sim 0.3$ for this example.     

In addition to the photon field of the companion, the blast wave also
encounters the diffuse emission of the stellar cluster. The latter is
dominated by that of the most massive stars in the cluster. For $N_*$
massive stars of luminosity $L_*$ each distributed isotropically
within a cluster of radius $R_{c}$, the ambient UV photon field in the
cluster is $U_{\rm UV}\sim 3N_*L_*/4\pi R_c^2c$. Fig.~\ref{fig:EC-all}
shows the total EIC emission coming for the scattering of the photon
field of the companion {\it and} the ambient cluster emission for
$N_*=30$, $L_*=10^{39.5}$ erg$\cdot$s$^{-1}$, and $R_{c}=5\times
10^{18}$ cm (thin lines).  Note that the ambient photon field is the
dominant source of soft photons for the more distant encounters (with
$\theta=10$, 20 deg) while it makes a modest contribution to the
$\theta=5$ deg example.  The peak of the EIC component is broader in
this case. Including both the companion and the diffuse sources of
soft photons, the total number of $\gamma$-ray photons emitted in the
0.1-300 GeV energy range increases to $N_{\gamma}\simeq 1.4\times
10^{52}$, $8\times 10^{51}$, $6\times 10^{51}$ for $\theta=5^{\rm
  o},10^{\rm o}, 20^{\rm o}$, respectively. In this case the emission
can be detected out to $d_p\simeq 3.3\times 10^{27}$ cm, or out to
$z\sim 0.3-0.4$.

\section{Discussion}
\label{sec:discussion}

If all, or at least the majority of, long duration GRBs come from the
death of massive stars, a fair fraction of them should take place
inside young, massive stellar clusters. In such crowded environments
the wind of the progenitor star terminates on sub-pc scales (mainly
due to interactions with other stellar winds). The wind-wind
interaction regions are characterized by shocks, contact
discontinuities and other sharp features in the density profile. The
blast wave, result of the GRB jet interacting with the external
medium, has a bumpy ride in such environments. In this work we search
for the characteristic observational signatures of such blast wave
evolution, focusing on the profile shaped by the collision of the wind
of the progenitor with that of a nearby massive star or `companion'.

\subsection{Optical and X-ray emission}
\label{sec:discuss_optxray}

As long as the Lorentz factor of the blast $\Gamma\gg 1$ and $\Gamma
\simmore\theta_j^{-1}$, one can study the basic features of the
blast-external medium interaction assuming spherical symmetry and
considering different lines of sight to the observer. We find that,
when the blast wave enters the shocked wind region, the optical light
curve at first flattens, followed by a steeper decline at later times
(Fig.~\ref{fig:opt}). When the characteristic synchrotron frequency
crosses the observed band, the modulations of the light curve are
enhanced. Qualitatively similar (but much less pronounced effects) are
in general expected in the X-rays as well (Fig.~\ref{fig:X-ray}). Our
results are in agreement with the generic arguments presented in
\citet{Nakar:2007yq}, who show that relativistic blast waves in
spherical symmetry do not show sharp features in their light curve
even for step-function changes of the external density as function of
radius.

Observations with dense sampling until late times and good
signal-to-noise are required to detect the relatively smooth changes
in the light curves. One recent example is a well studied optical,
X-ray afterglow of GRB 080413B \citep{Filgas11}, which shows a
characteristic flattening in the optical followed by a steeper decline
in both optical and X-ray wavelengths that appears to be in agreement
with the expectations from our model. Further examples include GRBs
021004 (see \cite{Lazzati02} and references therein), 080710
\citep{Kruehler09a}, 080129 \citep{Greiner09a} and 080913
\citep{Greiner09b}. Finally, there is an interesting GRB 071031
\citep{Kruehler09b}, which shows several bumps in the optical, which
might be due to more than one star influencing the afterglow (see,
however, the discussion in Sec.~\ref{sec:discuss_other}).

For a wide range of parameters, the blast wave is expected to enter
the shocked wind region when $\Gamma\sim \theta_j^{-1}$. Strictly
speaking, our 1D calculations are not directly applicable to this
regime since they ignore lateral spreading of the blast. Nevertheless,
recent 2D simulations performed by \cite{Zhang:2009fl} indicate that lateral
spreading effects are small at this stage and that the `jet breaks'
are mainly a geometric effect. We show that the combination of jet
collimation and structured external medium may lead to the rather sharp
features in the light curve, such as a fast decline of the light curve
as steep as $F\propto t^{-4...-5}$, which is not expected for the
evolution in a smooth external medium (Fig.~\ref{fig:opt-breaks}).

\subsection{$\gamma$-ray emission}
\label{sec:discuss_gammaray}

Another characteristic signature of the companion-blast interaction is
an inverse-Compton powered, GeV flare \citep{Giannios:2008ie}.  We
show that, under optimistic conditions, a large fraction of the energy
deposited on the shocked electrons is radiated away during the
encounter with the dense UV photon field of the massive star. Such a
flare at the energies of tens of GeV may account e.g., for the
emission of the GRB 940217 observed hours after the burst with {\it
  EGRET} \citep{Hurley94}. The {\it LAT} instrument on-board to the
{\it Fermi} satellite is capable of detecting such flaring for bursts
out to modest redshift $z\simless 1$. Depending of the distance of the
companion star, the $\sim$ GeV emission will peak on timescales that
range from a few hours up to a few weeks after the burst
(Fig.~\ref{fig:EC-all}). In some cases, the {\it ambient} UV photon
field of the massive stars in the cluster may be the dominant source
of the external inverse Compton component resulting in slower varying
$\gamma$-ray emission. A systematic search for delayed $\gamma$-ray
emission from the location of the burst on timescales of hours to
weeks after the burst may be fruitful. The timing and photon-energy of
such detection will provide invaluable information about the GRB
environment.

An additional promising source of seed photons to be inverse Compton
scattered at the forward shock may be the infrared emission from dust.
Though its physical origin is unclear, dust is forming actively in the
stellar (or shocked) wind region in many Wolf-Rayet stars (e.g.,
\citealt{Crowther03}). Up to $\sim10$\% of the stellar light can be
reprocessed into the infrared that provides an isotropic bath of
photons through which the blast propagates.  Mid-to-near infra red
photons can be scattered into multi TeV energies by the high-energy
tail of the electron distribution while still in the Thomson
scattering regime. Such emission, with a $\nu\cdot L_{\nu}$ peak at
$\sim$ TeV energies, may be detectable by atmospheric Cherenkov
telescopes provided that the burst takes place at redshift $z\simless
0.5$ (for the TeV photons to avoid attenuation while propagating
through the extragalactic background light).

\subsection{Other effects}
\label{sec:discuss_other}

In this paper, we focused on the effect of a \emph{single} massive
star on the circumburst medium density in the cluster where the GRB
takes place.  This star is most relevant in the early afterglow stages
since it lies the closest to the line of sight to the observer.
Clearly, the collective effect of many stellar winds will dominate the
density profile at larger distance (more than a few times the distance
to the `companion'). At these scales the external medium is expected
to be very inhomogeneous, although on scales much smaller than those
relevant for the blast wave emitting toward the observer. At
late times, therefore, the afterglow light curve may resemble that
expected from a constant density medium.  The confining effect of the
environment in the cluster can thus naturally limit the extend of the
wind of the progenitor and lead to a constant-like density as is
commonly suggested by modeling of afterglow observations (e.g., see
\citealt{Schulze11}).

\section*{Acknowledgments}
We are grateful to Miguel Angel Aloy, Jose Mar\'{\i}a Ib\'a\~{n}ez,
and Jochen Greiner for the constructive criticism and a fruitful
discussion. PM acknowledges the support from the European Research
Council (grant CAMAP-259276), from the Spanish Ministry of Education
and Science (AYA2007-67626-C03-01, AYA2010-21097-C03-01,
CSD2007-00050) and from the Valencian Conselleria d'Educaci\'o
(PROMETEO/2009/103). DG is a Lyman Spitzer, Jr Fellow of the
Department of Astrophysical Sciences of Princeton University. The
calculations have been performed on the
\href{http://www.uv.es/siuv/cas/zcalculo/calculouv/des_vives.wiki}{\emph{Llu\'{\i}s
    Vives}} cluster at the University of Valencia.

\bibliographystyle{mn2e}
\bibliography{colwinds}

\begin{thebibliography}{}

\bibitem[\protect\citeauthoryear{Blandford \& McKee}{Blandford \&
  McKee}{1976}]{Blandford:1976yg}
Blandford R.~D.,  McKee C.~F.,  1976, Physics of Fluids, 19, 1130

\bibitem[\protect\citeauthoryear{Chevalier \& Li}{Chevalier \&
  Li}{2000}]{Chevalier:2000aa}
Chevalier R.~A.,  Li Z.-Y.,  2000, \apj, 536, 195

\bibitem[\protect\citeauthoryear{Chevalier, Li \& Fransson}{Chevalier
  et~al.}{2004}]{CLF04}
Chevalier R.~A.,  Li Z.-Y.,  Fransson C.,  2004, \apj, 606, 369

\bibitem[\protect\citeauthoryear{Crowther}{Crowther}{2003}]{Crowther03}
Crowther P.~A.,  2003, Ap\&SS

\bibitem[\protect\citeauthoryear{Curran}{Curran et al.}{2003}]{Curran09}
Curran P.~A., Starling R.~L.~C., van der Horst A.~J., Wijers R.~A.~M.~J.,\ 2009, \mnras, 395, 580 

\bibitem[\protect\citeauthoryear{Figer}{Figer}{2004}]{Figer04}
Figer D.~F.,  2004, ASPC, 322, 49

\bibitem[\protect\citeauthoryear{{Filgas}}{{Filgas et al.}}{2011}]{Filgas11}
Filgas, R., et al.\ 2011, \aap, 526, A113 

\bibitem[\protect\citeauthoryear{{Frail}, {Kulkarni}, {Sari}, {Djorgovski},
  {Bloom}, {Galama}, {Reichart}, {Berger}, {Harrison}, {Price}, {Yost},
  {Diercks}, {Goodrich} \& {Chaffee}}{{Frail} et~al.}{2001}]{Frail01}
{Frail} D.~A. \etal,  2001, \apjl, 562, L55

\bibitem[\protect\citeauthoryear{Galama, Vreeswijk, van Paradijs, Kouveliotou,
  Augusteijn, B{\"o}hnhardt, Bre}{Galama et~al.}{1998}]{Galama98}
Galama T.~J., \etal, 1998, Nature 395, 670

\bibitem[\protect\citeauthoryear{Giannios}{Giannios}{2008}]{Giannios:2008ie}
Giannios D.,  2008, \aap, 488, L55

\bibitem[\protect\citeauthoryear{Greiner et
    al. 2009a}{Greiner et al.}{2009a}]{Greiner09a} Greiner, J., \etal, 2009a,
  \apj, 693, 1912

\bibitem[\protect\citeauthoryear{Greiner et
    al. 2009b}{Greiner et al.}{2009b}]{Greiner09b} Greiner, J., \etal, 2009b,
  \apj, 693, 1610

\bibitem[\protect\citeauthoryear{Hjorth, Sollerman, M{\o}ller, Fynbo, Woosley,
  Kouveliotou, Tanvir, Greiner}{Hjorth et~al.}{2003}]{Hjorth03}
Hjorth J., \etal,  2003, Nature, 423, 847

\bibitem[\protect\citeauthoryear{Hurley et al. 1994}{Hurley}{1994}]{Hurley94} 
Hurley K., \etal, 1994, Nature, 372, 652 

\bibitem[\protect\citeauthoryear{Kr\"uhler et
  al. 2009a}{Kr\"uhler et al.}{2009a}]{Kruehler09a} Kr\"uhler, T., \etal, 2009a, \aap, 508, 593

\bibitem[\protect\citeauthoryear{Kr\"uhler et
  al. 2009b}{Kr\"uhler et al.}{2009b}]{Kruehler09b} Kr\"uhler, T., \etal, 2009b, \apj, 697, 758

\bibitem[\protect\citeauthoryear{Lazzati et
  al. 2002}{Lazzati et al.}{2002}]{Lazzati02}
Lazzati D., Rossi E., Covino S., Ghisellini G., Malesani D.,\ 2002, \aap, 396, L5

\bibitem[\protect\citeauthoryear{Massey \& Hunter}{Massey \&
  Hunter}{1998}]{MH98}
Massey P.,  Hunter D.~A.,  1998, \apj, 493, 180

\bibitem[\protect\citeauthoryear{Mazzali et al. 2003}{Mazzali et al.}{2003}]{Mazzali03} 
Mazzali, P.~A., \etal, 2003, \apjl, 599, L95 

\bibitem[\protect\citeauthoryear{Mimica}{Mimica}{2004}]{Mimica:2004zz}
  Mimica P., 2004, Ph.D Thesis,
  \href{http://edoc.ub.uni-muenchen.de/archive/00002879}{LMU-M{\"u}nchen}
  

\bibitem[\protect\citeauthoryear{Mimica, Aloy \& M{\"u}ller}{Mimica
  et~al.}{2007}]{Mimica:2007aa}
Mimica P.,  Aloy M.~A., M{\"u}ller E.,  2007, \aap, 466, 93

\bibitem[\protect\citeauthoryear{Mimica, Aloy, Agudo, Marti, G{\'o}mez \&
  Miralles}{Mimica et~al.}{2009a}]{Mimica:2009aa}
Mimica P., Aloy M.~A., Agudo I., Mart\'{\i} J.~M., G\'omez J.~L.,
Miralles J. A.,  2009a, \apj, 696, 1142

\bibitem[\protect\citeauthoryear{Mimica, Giannios \& Aloy}{Mimica
  et~al.}{2009b}]{Mimica:2009bb}
Mimica P.,  Giannios D., Aloy M.~A.,  2009b, \aap, 494, 879

\bibitem[\protect\citeauthoryear{Mimica, Giannios \& Aloy}{Mimica
  et~al.}{2010}]{Mimica:2010cc}
Mimica P.,  Giannios D., Aloy M.~A.,  2010, \mnras, 407, 2501

\bibitem[\protect\citeauthoryear{Nakar \& Granot}{Nakar \&
  Granot}{2007}]{Nakar:2007yq}
Nakar E.,  Granot J.,  2007, \mnras, 380, 1744

\bibitem[\protect\citeauthoryear{{Panaitescu} \& {Kumar}}{{Panaitescu} \&
  {Kumar}}{2002}]{PK02}
{Panaitescu} A.,  {Kumar} P.,  2002, \apj, 571, 779

\bibitem[\protect\citeauthoryear{{Pe'er} \& {Wijers}}{{Pe'er} \&
  {Wijers}}{2006}]{PW06}
{Pe'er} A.,  {Wijers} R.~A.~M.~J.,  2006, \apj, 643, 1036

\bibitem[\protect\citeauthoryear{Piro, de Pasquale, Soffitta, Lazzati, Amati,
  Costa, Feroci, Frontera, Guidor}{Piro et~al.}{2005}]{Piro05}
Piro L., \etal,  2005, \apj, 623, 314

\bibitem[\protect\citeauthoryear{{Ramirez-Ruiz}, {Garc{\'{\i}}a-Segura},
  {Salmonson} \& {P{\'e}rez-Rend{\'o}n}}{{Ramirez-Ruiz} et~al.}{2005}]{RR05}
{Ramirez-Ruiz} E.,  {Garc{\'{\i}}a-Segura} G.,  {Salmonson} J.~D.,
  {P{\'e}rez-Rend{\'o}n} B.,  2005, \apj, 631, 435

\bibitem[\protect\citeauthoryear{Rhoads}{Rhoads}{1999}]{Rhoads:1999rt}
Rhoads J.~E.,  1999, \apj, 525, 737

\bibitem[\protect\citeauthoryear{Sari, Piran \& Narayan}{Sari
  et~al.}{1998}]{Sari:1998kx}
Sari R.,  Piran T.,  Narayan R.,  1998, \apjl, 494, L17

\bibitem[\protect\citeauthoryear{Schulze, Klose, Bj{\"o}rnsson, Jakobsson,
  Kann, Rossi, Kr{\"u}hler, Greiner}{Schulze et~al.}{2011}]{Schulze11}
Schulze S., \etal,  2010, \aap, 526, A23

\bibitem[\protect\citeauthoryear{Stanek, Matheson, Garnavich, Martini, Berlind,
  Caldwell, Challis, Brown, Sch}{Stanek et~al.}{2003}]{Stanek03}
Stanek K.~Z., \etal,  2003, \apj, 591, L17

\bibitem[\protect\citeauthoryear{Starling}{Starling
  et~al.}{2008}]{Starling08}
Starling R.~L.~C., van der Horst A.~J., Rol E., Wijers R.~A.~M.~J., 
Kouveliotou C., Wiersema K., Curran P.~A., Weltevrede P.,\ 2008, \apj, 672, 433 

\bibitem[\protect\citeauthoryear{Stevens, Blondin \& Pollock}{Stevens
  et~al.}{1992}]{Stevens:1992kx}
Stevens I.~R.,  Blondin J.~M.,    Pollock A. M.~T.,  1992, \apj, 386, 265

\bibitem[\protect\citeauthoryear{van Eerten, Meliani, Wijers \& Keppens}{van
  Eerten et~al.}{2009}]{vanEerten:2009em}
van Eerten H.~J., Meliani Z., Wijers R.~A.~M.~J., Keppens R,  2009, \mnras, 398, L63

\bibitem[\protect\citeauthoryear{van Marle, Keppens \& Meliani}{van Marle
  et~al.}{2011}]{vanMarle:2011jt}
van Marle A.~J.,  Keppens R., Meliani Z.,  2011, \aap, 527, A3

\bibitem[\protect\citeauthoryear{Wijers \& Galama}{Wijers \&
  Galama}{1999}]{Wijers:1999aa}
Wijers R. A. M.~J.,  Galama T.~J.,  1999, \apj, 523, 177

\bibitem[\protect\citeauthoryear{Wijers 2001}{Wijers}{2001}]{Wijers01} Wijers,
  R. A. M. J. 2001, Gamma Ray Bursts in the Afterglow Era,
  ed. E. Costa, F. Frontera, \& J. Hjorth (Springer: Berlin), 306

\bibitem[\protect\citeauthoryear{Wygoda, Waxman \& Frail}{Wygoda
  et~al.}{2011}]{Wygoda:2011wm}
Wygoda N.,  Waxman E., Frail D.,  2011, arXiv:1102.5618

\bibitem[\protect\citeauthoryear{Zhang \& MacFadyen}{Zhang \&
  MacFadyen}{2009}]{Zhang:2009fl}
Zhang W.,  MacFadyen A.,  2009, \apj, 698, 1261

\end{thebibliography}

\end{document}